\documentclass[twocolumn,prb]{revtex4}

\usepackage{epsfig,amssymb,amsmath}

\begin{document}

\title{Static charge-imbalance effects in intrinsic Josephson systems}

\author{J. Keller}
\author{D.A. Ryndyk}
\altaffiliation[On leave from ]{the Institute for Physics of Microstructures, RAS,
         Nizhny Novgorod, Russia}
\affiliation{Institut f\"ur Theoretische Physik, Universit\"at Regensburg, D-93040 Regensburg, Germany}

\begin{abstract}

Nonequilibrium effects created by stationary current injection in layered
d-wave superconductors forming a stack of intrinsic Josephson junctions are
studied. Starting from a nonequilibrium Green function theory we derive microscopic
expressions for the charge-imbalance (difference between electron- and
hole-like quasi-particles) on the superconducting layers and investigate its
influence on the quasi-particle current between the
layers. This nonequilibrium effect leads to shifts in the current-voltage
curves of the stack. The theory is applied to the interpretation of recent
current injection experiments in double-mesa structures.

Pacs: 74.72.-h, 74.50.+r, 74.40.+k.

\end{abstract}

\maketitle

\section{Introduction}

In the strongly anisotropic cuprate superconductors like
Bi$_2$Sr$_2$CaCu$_2$O$_{8+\delta}$ (BSCCO) the superconducting CuO$_2$ layers together
with the intermediate insulating material form a stack of Josephson junctions. In the
presence of a bias current perpendicular to the layers the intrinsic Josephson effect
manifests itself in a multibranch structure of the IV-curves,
\cite{Kleiner92,Kleiner94,Yurgens96,Yurgens00,Doh00} where on each branch a different
number of junctions is in the resistive state.

In the resistive state  high frequency Josephson oscillations at a finite
dc-voltage are accompanied by a dc-current which is carried mainly by unpaired
electrons (quasi-particles), while in the superconducting state it
is carried mainly by Cooper pairs. Therefore, on a superconducting layer
between a resistive and a superconducting junction the bias current
has to change its character from quasi-particle current to supercurrent. This
creates a non-equilibrium state on this layer with a finite quasi-particle
charge and a change of the condensate charge. The quasi-particle charge is
characterised by a distribution function with different numbers of
electronlike and holelike quasi-particles, while the change of the condensate
charge, the number of paired electrons, can be described by a shift of the
chemical potential of the condensate.

In recent experiments such non-equilibrium effects have been observed in layered
d-wave superconductors.\cite{Rother03} In a first type of experiments Shapiro steps
produced by high-frequency irradiation have been measured in mesa structures of BSCCO
with gold contacts. Here a shift  of the  step voltage from its canonical value
hf/(2e) was observed, which can be traced back   to a change of the contact resistance
due to quasi-particle charge on the first superconducting layer. In another type of
experiments current-voltage curves have been investigated for two mesas structured
close to each other on the same base crystal (see Fig.\,7 below). Here an influence of
the current through one mesa on the voltage drop through the other mesa has  been
measured which is caused by charge imbalance on the first common superconducting layer
of the base crystal. In a recent paper \cite{Ryndyk02} we have explained these effects
by using a semi-phenomenological approach based on  a microscopic non-equilibrium
Green function theory \cite{Ryndyk98, Ryndyk99} for layered superconductors.

In this paper we will present the full microscopic theory for stationary
charge-imbalance effects in intrinsic Josephson systems. In order to be specific, we
apply the theory first to an experiment, which in this form has not yet been done, but
is conceptual simpler then the double-mesa injection experiment mentioned above. We
consider a mesa as shown in Fig. 1 with two normal electrodes on top. Through one
electrode a stationary bias current is applied creating a charge imbalance on the
first superconducting layer. At the other electrode a voltage is measured as function
of the bias current through the first electrode with zero current through the second
electrode. This experiment is similar to the classical experiment by Clarke,
\cite{Clarke72} where a strong current injected into a bulk superconductor creates
quasi-particles. By two other electrodes, one normal junction and one Josephson
junction, a voltage is detected at zero current which measures the difference between
the chemical potential of quasi-particles and condensate. In our case the junction
between the normal electrode and the first superconducting layer is the normal
junction, the coupling between the first superconducting  layer and the next
superconducting layers is the Josephson junction. This experiment allows  to determine
the charge imbalance and (with further theoretical input) its relaxation rate. After
this investigation we return to the discussion of the double-mesa injection
experiment.

Nonequilibrium effects in {\it bulk-superconductors} have been studied a long time
ago. For an overview see the text-book by Tinkham \cite{Tinkham} and the
review-articles edited by Langenberg and Larkin.\cite{Langenberg} The basic concepts
of charge-imbalance have been developed by Tinkham and Clarke,\cite{Clarke72,
Clarke79} Pethick and Smith.\cite{Pethick79} The theory has been worked out using a
non-equilibrium Green function approach by Schmid and Sch\"on,
\cite{Schmid75,Schoen86}  and Larkin and Ovchinnikov.\cite{Larkin77,Larkin86}
Microscopic nonequilibrium theory of {\textit layered superconductors} was considered
first by Artemenko \cite{Artemenko80} and later by Graf et al. \cite{Graf95}

Nonequilibrium effects in {\textit intrinsic  Josephson system} have been
investigated theoretically already in various contexts using different methods
and  approximations.\cite{Koyama96,B96,Artemenko97,Preis98,Shafranjuk99,
Helm01,Helm00,Helm02,Bulaevskii02} In Refs. \onlinecite{Koyama96, Helm02} the
influence of charge-coupling on Josephson plasma oscillations has been
investigated, in the approach used in Refs. \cite{Artemenko97, Preis98,Helm01}
a systematic perturbation theory in the scalar potential on the layers has
been performed. In these theories charge imbalance is considered only
indirectly as far as it is induced by fluctuations of the scalar potential.
The present theory which uses non-equilibrium Green functions is more general.
Here charge imbalance is taken into account as an independent degree of
freedom, and therefore the results are different from those of earlier
treatments.

Here we apply this theory to stationary processes in layered
d-wave superconductors.  Due to the weak coupling between the
superconducting layers the total charge created by the bias-current is
non-zero.  Therefore we have to treat charge
fluctuations of the condensate and quasi-particles as independent quantities.
The d-wave character of the
order-parameter allows quasi-particle tunnelling also at low voltages and
temperatures. Furthermore it allows the relaxation of charge-imbalance by
elastic scattering processes (impurity scattering  within a layer). On the
other hand the weak coupling between the layers leads to a partial decoupling of
the kinetic equations for  the Green functions in different layers, making
this system a particularly simple example for the application of
nonequilibrium theory in superconductors.

In the following section we start with general considerations concerning the
Josephson effect in layered superconductors and the definition of
charge-imbalance. We then formulate the basic kinetic equations describing
stationary nonequilibrium effects in intrinsic Josephson systems and
calculate the charge imbalance microscopically. We show how these effects can be
measured by discussing the  above mentioned 4-point experiment and the
coupled two-mesa system. Finally we discuss the influence of
charge-imbalance on the IV-curves of a stack of junctions.

\begin{figure}
\begin{center}
\epsfxsize=0.7\hsize
\epsfig{figure=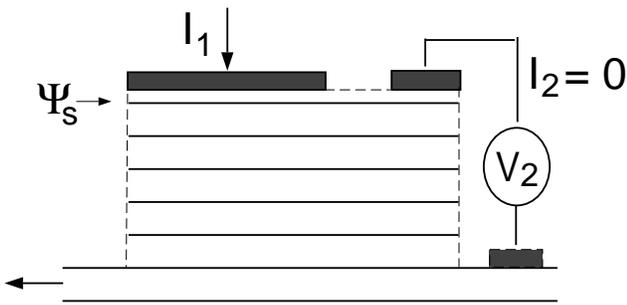,height=4cm}
\caption{Mesa-structure used for 4-point measurements consisting of
a stack of superconducting layers with two normal electrodes  on
top.}
\end{center}
\end{figure}

\section{Generalised Josephson equations}

We consider a system of superconducting layers with indices $n$ and order parameter
$\Delta_n(t) = \vert \Delta \vert \exp(i\chi_n(t))$ with time-dependent phase
$\chi_n(t)$. We define the gauge invariant phase difference as:
\begin{equation}
\gamma_{n,n+1}(t)= \chi_n(t) -
\chi_{n+1}(t) - \frac{2e}{\hbar} \int_n^{n+1} dz A_z(z,t) ,
\end{equation}
where $A_z(z,t)$ is the vector potential in the barrier. Here $e$
denotes the elementary charge. The charge of the electron is $-e$.

For the time
derivative of $\gamma_{n,n+1}$ we obtain the generalized Josephson relation:
\begin{equation}
\frac{d\gamma_{n,n+1}}{dt} = \frac{2e}{\hbar} \Bigl( V_{n,n+1} + \Phi_{n+1} -
\Phi_n\Bigr) . \label{genJos1}
\end{equation}
Here
\begin{equation}
V_{n,n+1}= \int_n^{n+1}dz E_z(z,t)
\end{equation}
is the voltage  and $\Phi_n(t)$ is the so-called
gauge invariant scalar potential defined by
\begin{equation}
\Phi_n(t)= \phi_n(t) - \frac{\hbar}{2e} \dot \chi_n(t),  \label{Phi}
\end{equation}
where $\phi_n(t)$ is the electrical scalar potential.

The quantity $\hbar \dot \gamma_{n,n+1} = 2e( V_{n,n+1} + \Phi_{n+1} - \Phi_n)$ is the
total energy required to transfer a Cooper pair from layer $n$ to $n+1$, $e\Phi_n$ can
be considered as the shift of the chemical potential of the superconducting condensate
with respect to an average chemical potential $\mu$, i.e. the number of particles in
the condensate is controlled by $\mu+e\Phi_n$. For equilibrium superconductors
$\hbar\dot \chi_n=2e\phi_n$, $\Phi_n=0$, and one has the usual Josephson relation
$\hbar\dot\gamma_{n,n+1}=2eV_{n,n+1}$.

The generalised Josephson relation (\ref{genJos1}) for a stack of Josephson junctions
can be written in another way showing explicitly the electric field distribution
between the layers. We split  the total charge fluctuation on a layer
$\delta\rho_n=\delta \rho^c_n + \delta \rho_n^q$ into a contribution $\delta \rho_n^c$
from particles in the condensate and a contribution $\delta \rho_n^q$ from unpaired
electrons (quasi-particles) describing charge-imbalance. The charge fluctuation of the
condensate can be expressed directly by the change of the chemical potential on layer
$n$:
\begin{equation}
\delta\rho^c_n= -2e^2N(0)\Phi_n,
\end{equation}
where $N(0)$ is the (two-dimensional) density of states for one  spin
direction at the Fermi energy. It is convenient to express also the
fluctuation of quasi-particle charge by some quasi-particle
potential $\Psi_n$, defining
\begin{equation}
\delta \rho^q_n = 2e^2N(0)\Psi_n.
\end{equation}
Then we obtain for the total charge density fluctuation:
\begin{equation}
\delta \rho_n = -2e^2N(0)(\Phi_n -\Psi_n) . \label{rho}
\end{equation}

While in bulk superconductors  charge neutrality leads to $\Phi=\Psi$,
this is  not the case in weakly coupled layered superconductors.

With help of (\ref{rho}) and the
Maxwell equation  ($d$ is the distance between the layers)
\begin{equation}
\delta \rho_n= \frac{\epsilon \epsilon_0}{d} (V_{n,n+1}-V_{n-1,n})
\label{Maxwell}
\end{equation}
the generalized Josephson relation now  reads:
\begin{align}\frac{\hbar}{2e }\dot \gamma_{n,n+1}& = (1+2 a)
V_{n,n+1} \nonumber \\
&- a (V_{n-1,n} + V_{n+1,n+2}) + \Psi_{n+1} - \Psi_n,
\label{genJos2}
\end{align}
with $a=\epsilon  \epsilon_0/(2e^2N(0)d)$. It shows that the Josephson
oscillation frequency is determined not only by the voltage  in the same
junction but also by the voltages  in neighboring junctions.
The coefficient $a$ has already been introduced by Koyama and
Tachiki,\cite{Koyama96} however in their theory charge imbalance effects
described by the quasi-particle potential $\Psi_n$ have not been considered.

For a barrier in the superconducting state the time average
$\langle\dot\gamma_{n,n+1}\rangle$ vanishes, while for a barrier in the resistive
state $\hbar\langle\dot\gamma_{n,n+1}\rangle/2$ is the electrochemical potential
difference. A similar equation also holds for the contact with the  normal electrode.
Denoting the normal electrode, which is assumed to be in equilibrium  by $n=0$, we
have
\begin{equation}
\frac{\hbar}{2e }\dot \gamma_{0,1} = V_{0,1}  + \Phi_1
 = (1+ a)
V_{0,1} + a V_{1,2} + \Psi_{1}.
\label{genJos3}
\end{equation}
In the case of a stack in contact with a normal electrode on top, where all internal
barriers are in the superconducting state we find for the first superconducting layer
$\Phi_1\simeq  aV+\Psi_1$. This result follows from Eqs. (\ref{genJos1},
\ref{genJos2}, \ref{genJos3}) in the limit $a\ll 1$. Note that the total voltage of
the stack is $V=\sum_n \dot \gamma_{n,n+1}\hbar/(2e)$.

In order to describe transport between the layers we need also an expression
for the current density. In our previous papers \cite{Ryndyk02,Rother03} we have used
\begin{align}
j_{n,n+1} &= j_c \sin\gamma_{n,n+1} \nonumber \\
&+ \frac{\sigma_{n,n+1}}{d} \Bigl( \frac{\hbar}{2e}
\dot \gamma_{n,n+1} + \Psi_n-\Psi_{n+1}\Bigr) ,
\label{current}
\end{align}
which  is approximately valid in the stationary case (no displacement current). The
first term is the current density of Cooper pairs. The rest is the quasi-particle
current density, which is driven not only by the electrochemical potential difference,
but contains a diffusion term proportional to the quasi-particle density difference.
In the following the current expression will be derived from the microscopic theory.

Finally we need some theory describing the creation and relaxation of
quasi-particle charge in order to calculate the charge-imbalance potential
$\Psi_n$. In our previous papers we have used a relaxation time
approximation. In the stationary case we obtained
\begin{equation}
\Psi_n= \tau_q(j^q_{n-1,n}-j^q_{n,n+1})/[2e^2N(0)] \label{rel}
\end{equation}
describing the balance between charge-imbalance creation by the in- and
outgoing quasi-particle currents and relaxation into thermal equilibrium inside
the layer. It will be the main task of this paper to derive such a relation
from a microscopic theory.

\section{Kinetic equations}

For the microscopic description we start from  equations of motion for the
spectral (
retarded and advanced) and Keldysh Green functions in Nambu space (for details see the
appendix\,\ref{A1}). Nonstationary corrections to the spectral functions are small,
and these functions can be taken in equilibrium. The Keldysh function contains the
necessary information about the nonequilibrium distribution functions. In our model we
consider d-wave superconductivity  on each layer within the BCS-approximation, elastic
impurity scattering inside the layer and tunnelling between neighboring layers in
lowest order. Then it is possible to formulate equations of motion for the Green
function $\hat G^K_n(\vec k,t_1,t_2)$ for each layer $n$ with a tunnelling self-energy
containing the coupling to the neighboring layers.

Furthermore we introduce quasiclassicle Green functions, which are obtained by
integrating the Green function over the kinetic energy $\xi_k=\epsilon_k-\mu$.
Performing in addition a  Fourier transformation with respect to the time difference
$\tau=t_1-t_2$ at fixed central time $t=(t_1+t_2)/2$ we obtain (see the
appendix\,\ref{A2}) the following kinetic equation for the quasiclassical Keldysh
function $\hat g^K_n(\hat k,t,\epsilon)$ in Nambu space:
\begin{align}
\left(\frac{i}{2}\frac{\partial}{ \partial t} + \epsilon\right) & \tau_3 \hat g^K
+\left(\frac{i}{ 2}\frac{\partial}{ \partial t} - \epsilon\right) \hat g^K \tau_3
\nonumber \\
&-\bigl\{\hat h(\hat k,t) \hat g^K-  \hat g^K\hat h(\hat k,t)\bigr\}
\nonumber \\ &
=\bigl\{\hat \sigma^R \hat g^K+ \hat \sigma^K\hat g^A -\hat g^R\hat \sigma^K - \hat
g^K \hat\sigma^A\bigr\}
\end{align}
depending on the central time $t$, the direction $\hat k$ of the momentum, and
the frequency $\epsilon$.
Here
\begin{align}
\hat h(\hat k,t) &= -e\Phi_n(t) - \hat \Delta(\hat k), \nonumber \\
\hat \Delta(\hat k) &=
\begin{pmatrix} 0 & \Delta(\hat k) \\ -\Delta(\hat k) & 0 \end{pmatrix},
\end{align}
where $\Phi_n(t)$ is the gauge invariant scalar potential Eq.(\ref{Phi}).
$\Delta(\hat k)$ is the superconducting order parameter with d-wave
symmetry. We have applied a gauge transformation such that on each layer
the superconducting order parameter has the same phase (in our case it is
real). The quantities $\hat \sigma$ are self-energies due to impurity
scattering and tunnelling, which will be specified later. The curly brackets
denote a convolution in time and frequency space. In the stationary case,
which we consider in the following, we will neglect the time dependence. Then
the convolutions are simple products, the dependence on $\Phi_n(t)$ drops
out, and we obtain the basic equation
\begin{align}
\epsilon \bigl( \tau_3 \hat g^K &- \hat g^K \tau_3 \bigr) + \hat \Delta \hat g^K- \hat
g^K\hat \Delta= \nonumber \\ &=\bigl[\hat \sigma^R \hat g^K+ \hat \sigma^K\hat g^A
-\hat g^R\hat \sigma^K - \hat g^K \hat\sigma^A\bigr].  \label{kineq}
\end{align}
This is actually an equation for the nonequilibrium distribution functions
contained in the Keldysh Green function $\hat g^K$ and the Keldysh
selfenergy $\hat \sigma_K$.

In the following we will
use the ansatz by Larkin and
Ovchinnikov:\cite{Larkin77, Larkin86}
\begin{equation}
\hat g^K = \bigl\{ \hat g^R  \hat d - \hat d \hat g^A \bigr\}
\end{equation}
with
$$
\hat d= \beta + \alpha \hat \tau_3
$$
containing  two distribution functions $\beta$ and $\alpha$.
In the stationary nonequili\-bri\-um case we can neglect the convolution and
replace it by  products:
\begin{equation}
\hat g^K= (\hat g^R - \hat g^A)\beta + (\hat g^R \tau_3 - \tau_3 \hat g^A) \alpha,
\label{LaOv}
\end{equation}
where for the retarded and advanced functions $\hat g^{R,A}(\epsilon)$ the
equilibrium functions can be used. In equilibrium
\begin{equation}
\beta(\epsilon)= \tanh(\epsilon/(2T)), \quad \alpha(\epsilon)=0.
\end{equation}

Returning now to the equation of motion Eq.(\ref{kineq}) and using
the traditional notation (see appendix A) for matrices in  Nambu space,
$g_{11}=:g, g_{12}=:f, g_{21}=:-f^+, g_{22}=:\bar g$,
the stationary part of the equations of motion for the normal and anomalous
Keldysh Green functions read explicitly
\begin{align}
-\Delta (f^{+K}- f^K) &= I_{11},  \label{11} \\
+\Delta (f^{+K}- f^K) &= I_{22},  \label{22} \\
2\epsilon f^K + \Delta (\bar g^K-g^K) &= I_{12},
\label{12} \\
2\epsilon f^{+K} + \Delta (\bar g^K-g^K) &= I_{21},
\label{21}
\end{align}
with the abbreviation for the tunnelling and scattering integrals: $$
I_{\alpha \beta} =\bigl[\hat \sigma^R \hat g^K+ \hat \sigma^K\hat g^A -\hat
g^R\hat \sigma^K - \hat g^K \hat \sigma^A\bigr]_{\alpha\beta}. $$

As we shell see later $\langle f \rangle=\langle \sigma_{12} \rangle=\langle \sigma_{21}
\rangle=0$ for d-wave symmetry, and we obtain:
\begin{align}
I_{11} &= \sigma^R_{11} g^{K} - g^R \sigma^K_{11} +
\sigma^K_{11} g^{A} - g^{K} \sigma^A_{11} \nonumber \\
I_{22} &= \sigma^R_{22} \bar g^{K} - \bar g^R \sigma^K_{22} +
\sigma^K_{22} \bar g^{A} - \bar g^{K} \sigma^A_{22} \nonumber \\
I_{12} &= \sigma^R_{11} f^K - f^R \sigma^K_{22}
 + \sigma^K_{11} f^A - f^K \sigma^A_{22} \nonumber \\
I_{21} &= -\sigma^R_{22} f^{+K} + f^{+R} \sigma^K_{11}
 - \sigma^K_{22} f^{+A} + f^{+K} \sigma^A_{11}
\end{align}

The scattering terms on the
r.h.s have different meaning: those which are proportional to the function
itself can be  considered as self-energy. Therefore we add these terms to the
l.h.s. of Eqs.(\ref{12}) and (\ref{21}), writing:
\begin{equation}
2\tilde \epsilon f^K + \Delta (\bar
g^K-g^K) =J_{12}
\end{equation}
with
\begin{equation}
\tilde \epsilon = \epsilon -  \frac{1}{2} (\sigma^R_{11}+\sigma^A_{11}) =
\epsilon - Re \sigma^R_{11}
\end{equation}
where we used the equilibrium values for $\sigma^{R,A}_{11}
=-\sigma^{R,A}_{22}$, and $\sigma^A_{11}= (\sigma^R_{11})^*$. Then the
correction to $\epsilon$ is just the real part of the scattering self-energy
evaluated below. The scattering term on the r.h.s. is then
\begin{align}
J_{12} &=
-f^R \sigma^K_{22} +\sigma^K_{11}f^A, \nonumber \\
J_{21} &=
f^{+R} \sigma^K_{11} -\sigma^K_{22}f^{+A}
\end{align}

Now we substitute  the anomalous Keldysh functions $f^K$ and $f^{+K}$ in
Eqs.({\ref{11}, \ref{22}) and obtain
\begin{equation}
I_{11}-I_{22} - \frac{\Delta }{ \tilde
\epsilon}\bigl(J_{12} - J_{21}\bigr)=0  \label{balance}
\end{equation}
\begin{equation}
I_{11}+I_{22}=0 \label{beta}
\end{equation}
These equations will be used in the following to determine the
distribution functions $\alpha(\epsilon)$ and $\beta(\epsilon)$ describing the
non-equilibrium state. In particular, Eq.(\ref{balance}) describes the balance
between the  relaxation of quasi-particle charge due to impurity scattering
within a layer and its creation by  the tunnelling
current (see Eq.(\ref{relax}) for the result in a special case).

\section{Calculation of the charge imbalance}

We consider potential scattering from randomly distributed
impurity centers and tunnelling to the neighboring layers in lowest order.
In order to calculate the self-energies and scattering integrals we need
expressions for the retarded and advanced Green functions. Here we use the
results from equilibrium (this can be justified by studying the
corresponding equations of motion in the low frequency limit).

\subsection{Retarded and advanced Green functions}

The retarded (advanced) quasi-classical Green functions in equilibrium have the form:
\begin{align}
g^R(\hat k,\epsilon)= \frac{\epsilon+i\gamma_\epsilon}{\sqrt{(\epsilon+
i\gamma_\epsilon)^2 - \Delta^2(\hat k)}}, \quad g^A = - (g^R)^* , \nonumber \\
f^R(\hat k,\epsilon)= \frac{\Delta(\hat k)}{\sqrt{(\epsilon+ i\gamma_\epsilon)^2 -
\Delta^2(\hat k)}}, \quad f^A = - (f^R)^* ,
\end{align}
with $-i\gamma_\epsilon= \sigma_{11}^R(\epsilon)$.
Here we use the sign-convention $Im \sqrt{\phantom x} >0$.
Furthermore we have for the equilibrium functions $f^{+R,A}= f^{R,A}$, $\bar
g^{R,A}= - g^{R,A}$. In the following we will also need the
combinations:
\begin{align}
u(\hat k,\epsilon)&=\frac{1}{2}\bigl(g^R(\hat k,\epsilon)
- g^A(\hat k,\epsilon)\bigr) , \nonumber
\\ v(\hat k,\epsilon)&=\frac{1}{2}\bigl(f^R(\hat k,\epsilon) - f^A(\hat
k,\epsilon)\bigr) , \nonumber \\ w(\hat k,\epsilon)&=\frac{i}{2}\bigl(f^R(\hat
k,\epsilon) + f^A(\hat k,\epsilon)\bigr) .
\end{align}
The (even) spectral function $u(\hat k,\epsilon)$ is the tunnelling density of
states. With this notation we obtain for the Keldysh functions Eq.(\ref{LaOv})
\begin{align}
g^K(\hat k,\epsilon) &=2u(\hat k,\epsilon)(\alpha(\hat k,\epsilon) + \beta(\hat
k,\epsilon)), \nonumber \\
\bar g^K(\hat k,\epsilon) &= 2u(\hat k,\epsilon)(-\beta(\hat
k,\epsilon)+\alpha(\hat k,\epsilon)) .
\end{align}

\subsection{Potential scattering}

The
self-energies in Born approximation are given by:
\begin{equation}
\hat \sigma_p^{R,A,K}(\epsilon) = c \sum_{k'}\vert V_0\vert^2 \hat G^{R,A,K}(\vec
k,\epsilon)= -i\nu_p\langle \hat g^{R,A,K}(\epsilon) \rangle
\end{equation}
with $\nu_p=c\pi N(0)\vert V_0\vert^2$. Here $V_0$ is the scattering
potential and $c$ the impurity concentration. For comparison in the numerical
results we also use the t-matrix approximation in the strong scattering limit.
This is obtained by replacing $\nu_p$ by $\nu_p/\vert g^R(\epsilon)\vert^2$.

For the calculation of the Keldysh component of the
self-energy we use the ansatz by Larkin and
Ovchinnikov Eq.(\ref{LaOv}) and obtain
\begin{equation}
\sigma_p^K(\epsilon)= - 2i\nu_p \langle u(\epsilon) \beta(\epsilon)\rangle \tau_3 -
2i\nu_p \langle v(\epsilon) \alpha(\epsilon)\rangle \tau_0 .
\end{equation}

Using these results we obtain for the potential part of the scattering
terms:
\begin{align} \label{potscatt}
I^p_{11}=-I^p_{22}&= 4i\nu_p\bigl[u(\epsilon)\langle
u(\epsilon)\alpha(\epsilon)\rangle -\langle u(\epsilon)\rangle u(\epsilon)
\alpha(\epsilon)\bigr] , \nonumber \\
J^p_{12}- J^p_{21}&= 8i\nu_p v(\epsilon) \langle
u(\epsilon)\alpha(\epsilon,t)\rangle . \end{align}

\subsection{Tunneling}

We describe the coupling between neighboring layers by the usual
tunnelling Hamiltonian
\begin{equation}
H=\sum_{n\sigma k k'} (t_{k'k}e^{-\frac{2e}{\hbar}\int_n^{n+1} dz A_z(z,t)}
c^\dagger_{n+1,k'\sigma} c_{n,k\sigma} + h.c.) ,\end{equation}
which after the gauge transformation becomes
\begin{equation}
H=\sum_{n\sigma k k'} (t_{k'k}e^{i\gamma_{n,n'}(t)}c^\dagger_{n+1,k'\sigma}
c_{n,k\sigma} + h.c.). \end{equation}
Then we obtain for the
contribution to the tunnelling self-energy from the coupling between layers
$n$ and $n'=n\pm 1$:
\begin{equation}
\begin{array}{c}\displaystyle
\hat \sigma_t(\epsilon,t)_{n,n'}= \int d(t_1-t_2)e^{i\epsilon(t_1-t_2)} \sum_{k'}
\vert t_{k'k}\vert^2 \\ \displaystyle
e^{-i\tau_3\gamma_{n,n'}(t_1)/2}\hat G(\vec k',t_1,t_2)e^{+i\tau_3\gamma_{n,n'}(t_2)/2
} .
\end{array}
\end{equation}
If we neglect additional phase fluctuation, the phase
difference is $\gamma_{n,n'}(t)= \gamma^0_{n,n'} + \Omega_{n,n'} t$, where
$\gamma^0_{n,n'}$ is the static phase difference, which is determined by the
supercurrent, and $\Omega_{n,n'}$ is proportional to  the electrochemical
potential difference between the two layers.

For the tunnelling matrix element we have to
assume a partial conservation of momentum in order to obtain Josephson
coupling between different superconducting layers with d-wave order parameter.
This, however, is  not relevant for the tunnelling of quasi-particles, which
will be discussed here primarily. In the following we will model the
tunnelling matrixelement near the Fermi surface by $\pi N(0) \vert
t_{k'k}\vert^2 = \nu_t(\hat k,\hat k')$ with $\langle \Delta(\hat k)\nu_t(\hat
k, \hat k')\Delta(\hat k')\rangle \ne 0$.

The most simple case to study nonequilibrium effects is the tunnelling
between a normal electrode and a superconducting layer, which will be
discussed in the following. Results for quasi-particle tunnelling
between two superconducting layers will be given in the appendix.
In the case of tunnelling with a normal elctrode only
the normal Green functions contribute to the tunnelling
self-energy, which is time-independent. Writing $v= \Omega_{n,n'}/2$ we obtain
\begin{align}
\hat\sigma_t^R(\hat k,\epsilon)_{(n,n')}= &-\frac{i}{2} \Bigl(\langle
\nu_t{g}^R(\epsilon-v)\rangle'
- \langle \nu_t{g}^R(\epsilon+v)\rangle'\Bigr)\tau_0 \nonumber \\
-\frac{i}{ 2} \Bigl(\langle \nu_t{g}^R(\epsilon&-v)\rangle'
+ \langle \nu_t{g}^R(\epsilon+v)\rangle'\Bigr)\tau_3 ,
\end{align}
 \begin{align}
&\hat\sigma_t^K(\hat k,\epsilon)_{(n,n')}= \\
&-i \Bigl(\langle \nu_t u'(\epsilon-v)\beta'(\epsilon-v)\rangle'
-\langle \nu_t u'(\epsilon+v)\beta'(\epsilon+v)\rangle' \Bigr)\tau_0
\nonumber \\
&-i\Bigl(\langle \nu_t u'(\epsilon-v)\beta'(\epsilon-v)\rangle'
+ \langle\nu_t u'(\epsilon+v)\beta'(\epsilon+v)\rangle' \Bigr)\tau_3
.\nonumber
\end{align}
The prime denotes spectral functions and distribution functions on layer
$n'$ and $\langle \nu_t {g}^R\rangle'= \langle \nu_t(\hat k,\hat k')g^R(\hat
k',\epsilon)\rangle_{k'}$ denotes an average over the direction of $\hat k'$.

These results will be used to calculate the tunnelling contributions
to the scattering terms (\ref{balance},\ref{beta}) in the kinetic equation.
Denoting the superconducting layer by $n$, the normal layer by $n'$, replacing
the spectral function on the normal layer by $u'(\epsilon)=1$, and neglecting
terms of order $\nu_t \alpha$ we find
\begin{align} \label{tunnscatt}
I^{tn}_{11}-
I^{tn}_{22} &=
4i\langle \nu_t(\beta'(\epsilon-v ) -
\beta'(\epsilon+v))u(\epsilon)\rangle' , \nonumber \\
I^{tn}_{11}+
I^{tn}_{22} &=4i\langle \nu_t(\beta'(\epsilon-v ) +
\beta'(\epsilon+v)-\beta(\epsilon)u(\epsilon)\rangle' ,\nonumber
\\
J^{tn}_{12}-
J^{tn}_{21}
&= 4i\langle \nu_tv(\epsilon)(\beta'(\epsilon-v)-\beta'(\epsilon+v ))
\rangle' .
\end{align}

In order to determine the charge-imbalance on the superconducting layer in
the stack we also need the quasi-particle contribution from tunnelling into
the neighboring superconducting layer with the  barrier in the superconducting
state. As will be shown in the Appendix, this contribution vanishes if we
neglect terms of order $\nu_t \alpha$.

\subsection{Solution of the kinetic equation}

Inserting now the different contributions to the scattering term from
potential scattering Eq.(\ref{potscatt}) and tunnelling Eq.(\ref{tunnscatt})
into Eq.(\ref{balance}) we obtain the
following  equation determining the distribution function $\alpha(\hat
k,\epsilon)$ on the superconducting layer:
\begin{align}
2\nu_p\bigl[u(\epsilon)\langle u(\epsilon)\alpha(\epsilon)\rangle
&-\langle u(\epsilon)\rangle u(\epsilon) \alpha(\epsilon)\bigr] \\
&- 2\nu_p\frac{\Delta
v(\epsilon)}
{\tilde \epsilon} \langle u(\epsilon)\alpha(\epsilon)\rangle \nonumber \\
= - \bigl(u(\epsilon)- \frac{\Delta}{ \tilde
\epsilon}v(\epsilon)\bigr)
&\bigl\langle \nu_t(\hat k,\hat k') \bigl[
\beta'(\epsilon-v) - \beta'(\epsilon+v))\bigr] \bigr\rangle' .\nonumber
\end{align}
The r.h.s. of this equation describes charge imbalance generation by
tunnelling of quasi particles. The l.h.s. describes charge imbalance
relaxation due to impurity scattering.
Taking an angular average and defining
\begin{align} \tilde \alpha(\epsilon) &:=\langle
u(\epsilon)\alpha(\epsilon)\rangle,  \nonumber \\
R(\epsilon)&:= u(\epsilon)- \frac{\Delta v(\epsilon)}{ \tilde
\epsilon}, \nonumber \\
\nu(\epsilon)&:=2\nu_p\langle \frac{\Delta v(\epsilon)}{ \tilde
\epsilon}\rangle ,
\end{align}
we obtain
\begin{equation}
\tilde \alpha(\epsilon)= \frac{\langle \nu_t
R(\epsilon)\rangle }{\nu(\epsilon)}
(\beta'(\epsilon-v)-\beta'(\epsilon+v)). \label{relax}
\end{equation}

Finally from Eq.(\ref{beta}) and Eq.(\ref{tunnscatt}) we obtain for the
distribution function on the superconducting layer
\begin{equation}
\beta(\epsilon)= \frac{1}{2}(\beta'(\epsilon-v) + \beta'(\epsilon+
v)) ,
\end{equation}
where $\beta'(\epsilon)$ is the distribution function on the normal layer,
which is assumed to be in equilibrium. This simple relation is only true as
long as we neglect inelastic scattering, which is necessary for a relaxation
of quasi-particles into the condensate.

\subsection{Charge-imbalance}

Equation (\ref{relax}) is the main result of the paper.
The function $\tilde \alpha(\epsilon)$ describes the charge of
quasi-particles with energy
$\epsilon$ on the superconducting layer. It is
related to the quasi-particle potential $\Psi$ introduced  in the
phenomenological theory in the following way:
The quasiclassical expression for the charge density is
\begin{equation}
\delta\rho_n=-2eN(0)\left[e\Phi_n+\int_{-\infty}^{\infty}\frac{d\epsilon}{4}
\left\langle g^K_n(\hat k,\epsilon)\right\rangle\right].
\end{equation}
Using the distribution function $\alpha$, introduced in the previous section, we can write
this expression as
\begin{align}\label{rho_kin-d}
\delta\rho_n &=-2eN(0)\left[e\Phi_n-\int_0^{\infty}d\epsilon
\left\langle u(\hat k,\epsilon)\alpha(\hat k,\epsilon)\right\rangle\right]
\nonumber \\
 &=-2e^2N(0)[\Phi_n-\Psi_n],
\end{align}
with the charge-imbalance potential determined by the formula
\begin{equation}\label{Psi}
  \Psi=(1/e)\int_0^{\infty}d\epsilon\langle u(\hat k,\epsilon)\alpha(\hat k,\epsilon)\rangle.
\end{equation}

Relation of $\alpha(\hat k,\epsilon)$ to the "clean limit" charge-imbalance
distribution function $\alpha_k$ in k-space is given in Appendix D.
The relaxation rate
$\nu(\epsilon)$  describes the relaxation into thermal equilibrium of the
difference between electron and hole-like quasi-particles due to impurity
scattering. It replaces the relaxation rate $1/\tau_q$ in Eq.(\ref{rel}).
The r.h.s. of Eq.(\ref{relax}) contains the  tunnelling of quasi-particle
charge from the normal electrode into the superconducting layer due to the
applied voltage. It is similar but not equal to the tunnelling current at the
same energy $\epsilon$. In the tunnelling current (see below) the function
$R(\epsilon)$ is replaced by the density-of-states function  $u(\epsilon)$. In
the absence of impurity scattering one finds $R(\epsilon)= 1/u(\epsilon)$ in
agreement with Ref.\onlinecite{Clarke79}.

\section{Calculation of the tunnelling current}

In order to measure the charge-imbalance induced by the quasi-particle
injection we  also need an expression for the
current, in particular between the normal contact and the first
superconducting layer. Quite  generally the current between neighboring
layers $n$ and $n'$ in lowest order in the tunnelling matrix element can be
written as
\begin{align}
&J_{n,n'}(t) = + \frac{2e}{\hbar}\sum_{ kk'} \int dt_1 \vert t_{kk'}\vert^2
\cr
\Bigl(& \hat G^R_n(k,t,t_1)
e^{-i\tau_3\gamma_{n,n'}(t_1)/2}
\hat G^<_{n'}(k,t_1,t)e^{+i\tau_3\gamma_{n,n+1}(t)/2}\nonumber
\\ & \hat G^<_n(k,t,t_1)
e^{-i\tau_3\gamma_{n,n'}(t_1)/2}\hat G^A_{n'}(k,t_1,t)
e^{+i\tau_3\gamma_{n,n'}(t)/2}\Bigr)_{11} \nonumber \\
&+ c.c .
\end{align}
 In the
stationary case (constant applied voltage) we neglect the time dependence of
the Green functions on the central time and use a Fourier
transformation with respect to the time difference as defined
above. Furthermore we restrict ourselves to the tunnelling between the
normal layer and the first superconducting layer. Then only the normal Green
functions contribute and we obtain (with $v=\Omega_{n,n'}/2$)
the usual expression for the
quasi-particle current density
\begin{align}
j_{n,n'}(v) &= - \frac{e}{\hbar}  N(0)  \int_{-\infty}^{+\infty}
{d\epsilon} \nonumber \\
\Bigl( &\bigl\langle\negthickspace\bigl\langle \nu_t(\hat k,\hat k') g^R(\hat k,\epsilon)
g^<(\hat k',\epsilon-v)\rangle \nonumber \\ &+ \langle \nu_t(\hat k,\hat
k')g^<(\hat k,\epsilon) g^A(\hat
k',\epsilon-v)\bigr\rangle\negmedspace\bigr\rangle \Bigr) +
c.c .
\end{align}
Here the prime denotes momenta and distribution functions on layer $n'$, and
the double brackets denote an average over both $\hat k$ and $\hat k'$.

Now we express the lesser functions $g^<$ by the Keldysh functions
\begin{equation}
\hat g^<= \frac{1}{ 2}( \hat g^K - \hat g^R- \hat g^A)
\end{equation}
and express the latter by the non-equilibrium distribution functions $\beta$
and $\alpha$, then
\begin{align}
j_{n,n'}(v) &= - \frac{2e}{\hbar} N(0)  \int_{-\infty}^{+\infty} {d\epsilon}
\bigl\langle\negthickspace\bigl\langle\nu(\hat k,\hat k')
u(\hat k,\epsilon)u'(\hat
k',\epsilon-v)\nonumber \\
&\quad \quad \bigl[\beta'(\epsilon-v)-\beta(\epsilon)+
\alpha'(\epsilon-v)-\alpha(\epsilon)\bigr]
\bigr\rangle\negthinspace\bigr\rangle \nonumber \\
\label{current2}
&=:j_{n,n'}^\beta(v)+ j^\alpha_{n,n'}(v) .
\end{align}
This is another important result. The current between neighboring layers $n$
and $n'$ is the sum of two parts containing the distribution functions
$\beta(\epsilon)$ and $\alpha(\epsilon)$ respectively. The current $j^\beta(v)$ is the
quasi-particle current driven and created by the electro-chemical potential
difference between the two layers. The current $j^\alpha(v)$ describes the
diffusion current driven by the charge imbalance. Both current contributions
depend on the density of state $u(\epsilon)$ of the two layers.

For the further application to the tunnelling between a normal electrode and a
superconducting layer it is convenient to denote by $j$ the current flow
{\it from} the normal electrode {\it to} the superconducting layer and by
$V$ the voltage drop in this direction (i.e. $v=-eV$). With $u'(\epsilon)=1$,
$\alpha'=0$, $\beta'(\epsilon)=\beta_0(\epsilon)=\tanh(\epsilon/(k_BT))$
for the normal layer, and exploiting the fact that $u(\epsilon)$ and
$\alpha(\epsilon)$ are even functions, while $\beta(\epsilon)$ is an odd
function of $\epsilon$ we obtain for the current contributions:
\begin{align}
j^\beta(V) =  \frac{e }{\hbar}N(0) \nu_t&
\int_{-\infty}^{+\infty} {d\epsilon} \langle u(\hat k,\epsilon)\rangle
\bigl[\beta_0(\epsilon+eV) \nonumber \\
&-\beta_0(\epsilon-eV)\bigr]=:j_0(V) ,
\label{current3} \\
j^\alpha(V)= - \frac{2e }{\hbar}N(0) \nu_t&
\int_{-\infty}^{+\infty} {d\epsilon} \langle u(\hat
k,\epsilon)\alpha(\hat k,\epsilon)\rangle \nonumber \\
=: -\sigma_0 \Psi(V)& .\label{current4}
\end{align}
Here we have replaced  the tunnelling rate by
some  average  $\nu_t=\langle \nu(\hat k,\hat k')\rangle $ and
have defined the ohmic resistance $\sigma_0=4e^2N(0)\nu_t/\hbar$.
Then the current driven by the nonequilibrium distribution of
quasi-particles can be expressed by the quasi-particle potential
\begin{align}\label{imbalance}
\Psi(V) &= \frac{1}{e}\int_0^\infty d\epsilon
\langle u(\hat k,\epsilon)\alpha(\hat k,\epsilon)\rangle)  \\
&=\frac{1}{e}\int_0^\infty d\epsilon \frac{\nu_t}{\nu(\epsilon)}
\langle  R(\epsilon)\rangle[\beta_0(\epsilon+eV) -\beta_0(\epsilon-eV)]
.\nonumber
\end{align}

\section{Results}

\begin{figure}
\begin{center}
\epsfxsize=0.8\hsize
\epsfig{figure=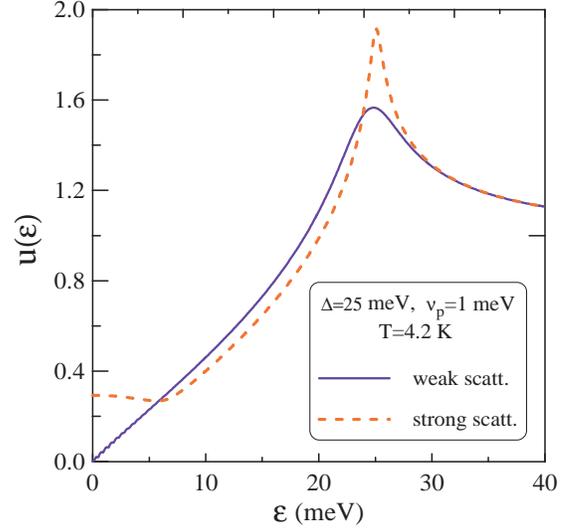,height=7cm}
\caption{Spectral function (tunnelling density of states) calculated for a
d-wave superconductor with impurity scattering.}
\end{center}
\end{figure}

For our model calculations we use a d-wave order parameter with
an angular dependence  of the usual form $\Delta(\hat k)=
\Delta(T)\phi(\hat k)$ with $\phi(\hat k)=\cos 2\theta_k$.
The tunnelling matrix element will be
parametrised as $\nu_t(\hat k,\hat k')= \nu_1 + \nu_2 \phi(\hat k)\phi(\hat k')$.
This simplifies the averaging procedure: the parameter $\nu_1$ enters the normal
tunnelling probability while the parameter $\nu_2$ determines the Josephson
coupling.

\subsection{Tunneling current and charge imbalance}

In Fig.\,2 we show well-known typical results for the spectral function (tunnelling density
of states) $u(\epsilon)$ at low temperatures with a self-consistently
determined  self-energy $ i\gamma(\epsilon)$ for the two limiting
cases of Born scattering and in the unitary limit. Note that in the unitary
limit the spectral function stays finite for $\epsilon\to 0$. This function
will be needed as input for the following calculations of the
tunnelling current and the charge-imbalance relaxation.
\begin{figure}
\begin{center}
\epsfxsize=0.8\hsize
\epsfig{figure=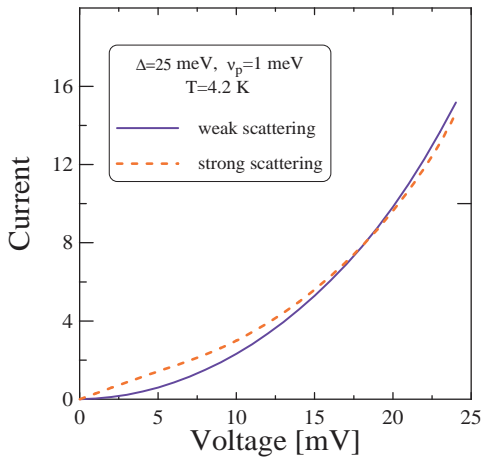,height=6cm}
\caption{Tunnelling current $j_0/\sigma_0$[mV] as function of the applied voltage.}
\end{center}
\end{figure}
Fig.\,3 shows the corresponding normalized
 tunnelling current $j_0(V)/\sigma_0$ calculated from Eq.(\ref{current3}) between a
normal electrode and a superconducting layer as function of the voltage-drop
$V$  between the normal electrode and the superconducting layer.
\begin{figure} \begin{center}
\epsfxsize=0.8\hsize
\epsfig{figure=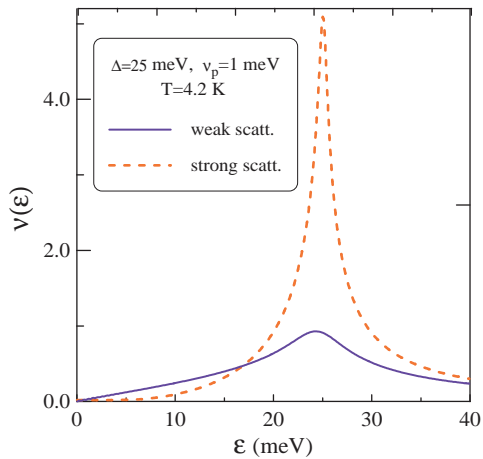,height=6cm}
\caption{Relaxation rate for the charge-imbalance due to impurity scattering as function of
energy.}
\end{center}
\end{figure}

Now we  turn to the calculation of charge imbalance.
In Fig.\,4 the frequency dependence of the relaxation function
$\nu(\epsilon)$ is shown for low temperatures. The frequency dependence
reflects  the available phase-space for elastic scattering
processes from electron-like into hole-like quasi-particles at the energy
$\epsilon$.  With help of this function we  calculate from
Eq.(\ref{imbalance}) the charge imbalance potential $\Psi$ generated on a superconducting layer in contact
with a normal layer. Fig. 5a shows $\Psi(V)$  as function
of the voltage $V$ between the normal electrode and the superconducting
layer.



\begin{figure}
\begin{minipage}{7cm}
\epsfxsize=0.8\hsize
\epsfig{figure=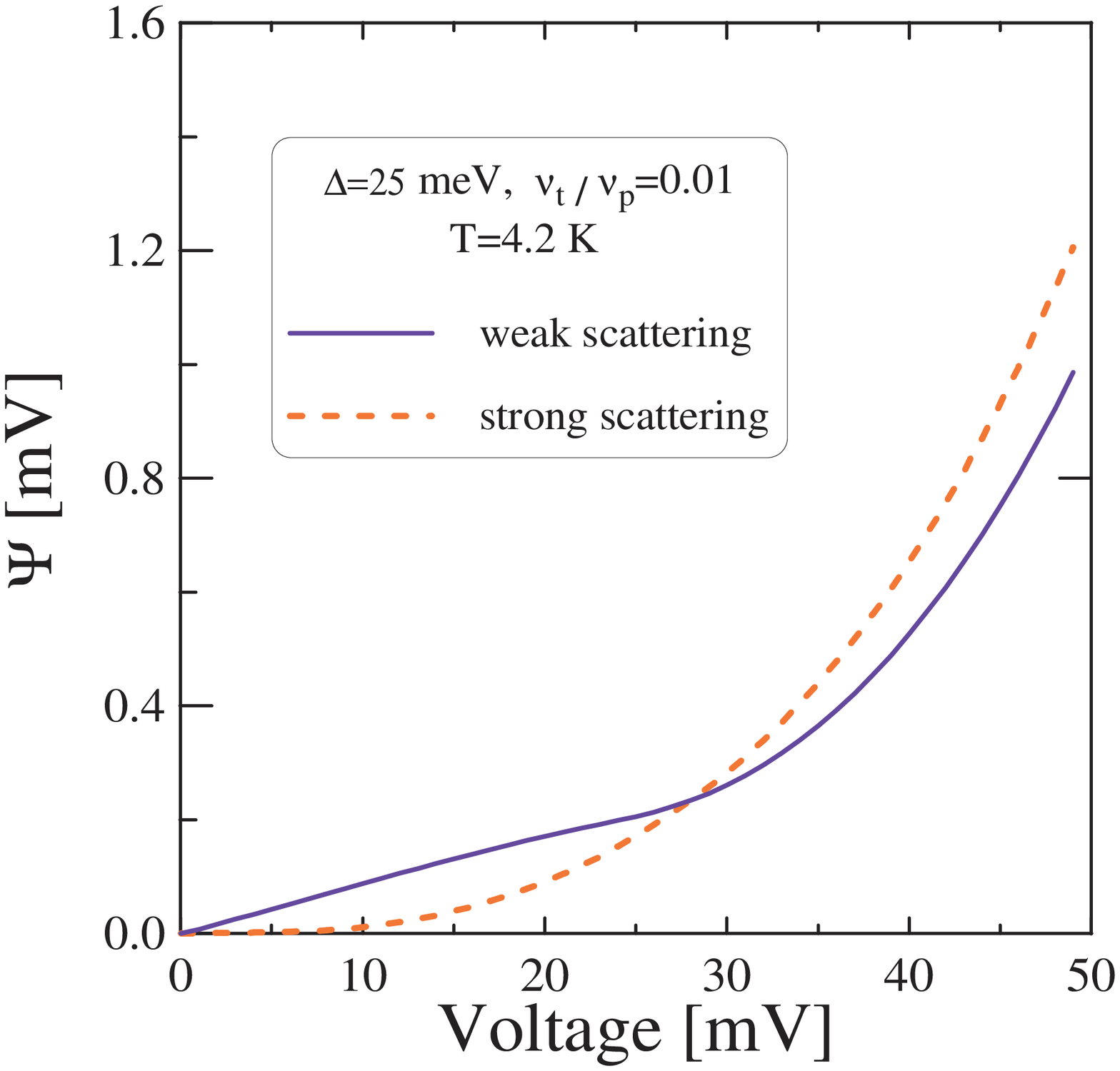,height=6cm}
\epsfig{figure=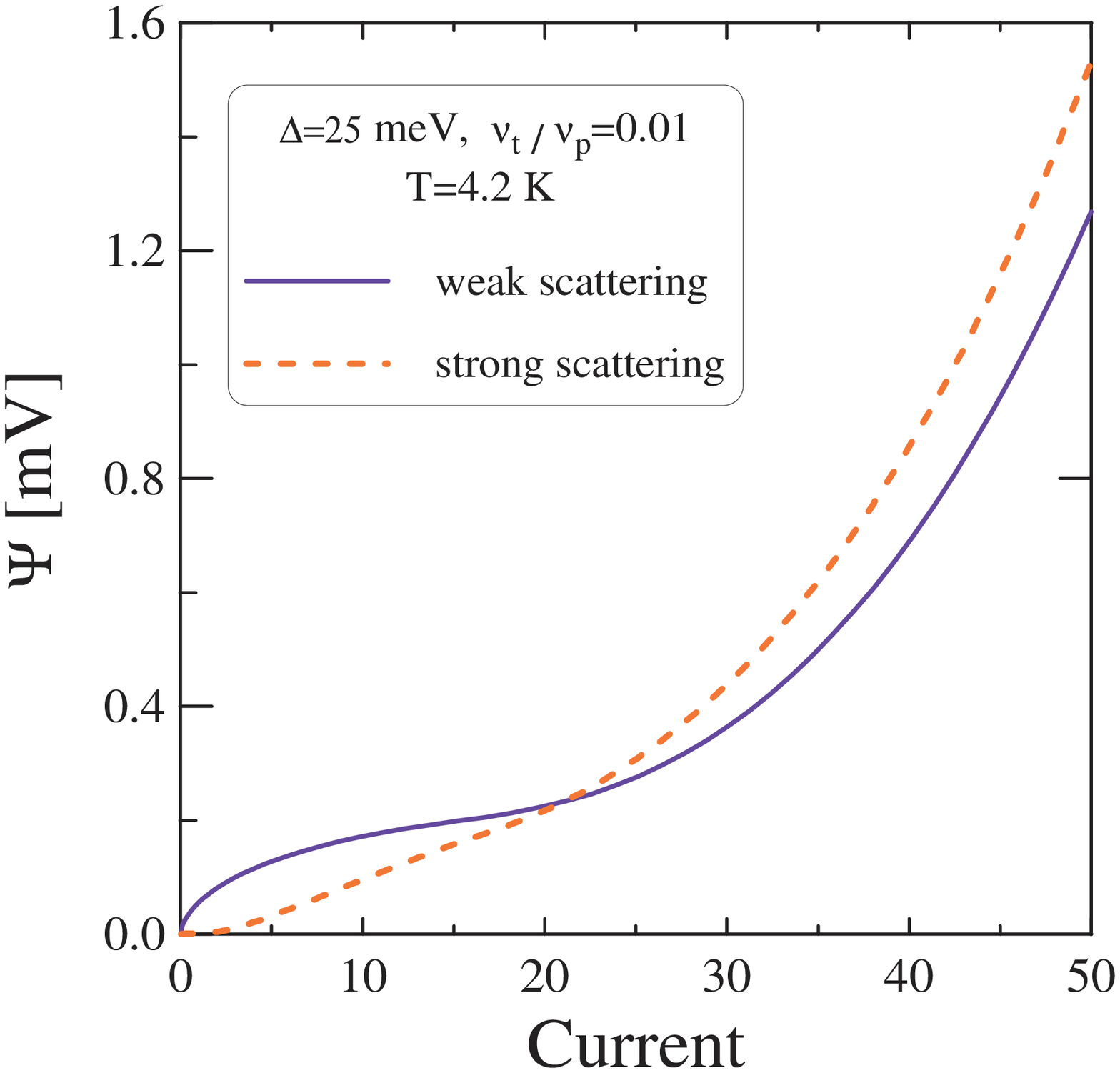,height=6cm}
\end{minipage}
\caption{Charge imbalance potential generated on a superconducting layer in
contact with a normal electrode as function of the applied voltage (a) and
the tunnelling current $j_0/\sigma_0$ (b)}
\end{figure}

As $V$ cannot easily be measured directly, we express it by the
corresponding current density.
The resulting function $\psi(j)$ defined by $\psi(j)=\Psi(V)$ with $j=j_0(V)$
given by Eq.(\ref{current3})
is shown in Fig. 5b. It depends on the ratio $\nu_t/\nu_p$ between the
average tunnelling rate and the potential scattering rate. The nonlinear dependence on
the current reflects to some extent the nonlinear current voltage curve.

\subsection{Experiment with two normal electrodes}

Now we apply the theory to the basic experiment, where one superconducting
layer is in contact with two normal electrodes. Through the first
electrode  with area $F_1$ a
current $I_1$ is applied, which creates a charge imbalance $\Psi_s$ on the
superconducting layer. At a second electrode the voltage $V_2$ is measured
with no current flowing.
We assume that charge imbalance spreads evenly over the whole
superconducting layer of size $F$.
Then the  charge imbalance potential created on the superconducting layer is
given by $\Psi_s = (F_1/F) \psi(j_1)$,
where $j_1=I_1/F_1$ is the current density  through the electrode (1), with
$\psi(j)$ defined above and shown in Fig.\,5b.

In order to determine the voltage $V_2$ measured at the second electrode
we use  the current equations (\ref{current3}, \ref{current4}) and exploit the
condition that no current is flowing, $j_2=j_0(V_2)-\sigma_0
\Psi_s=0$, i.e. we have a compensation of the
quasi-particle current driven by the voltage and the quasi-particle diffusion
current driven by the charge imbalance potential.
Using for $\Psi_s$ the value determined above we find
\begin{equation}
\int_0^\infty d\epsilon \langle u(\hat k,\epsilon )
[\beta_0(\epsilon+eV_2) - \beta_0(\epsilon-eV_2)] = \frac{F_1}{F}
\psi(j_1) .
\end{equation}
Thus we obtain
$V_2$  as function of $j_1$. Results are shown in Fig. 6 for the
case of a very small test electrode, $F_2\ll F$, $F_1\simeq F$.
Note that in this channel the
electrochemical potential drop is only between the normal electrode and the
first superconducting layer, all the other barriers in the stack have
$\Omega_{n,n+1}=0$. Furthermore, if the electrodes are in equilibrium,
the total voltage equals the total
electrochemical potential, thus $V_2$ is the total measured voltage.
\begin{figure}
\begin{center}
\epsfxsize=0.7\hsize
\epsfig{figure=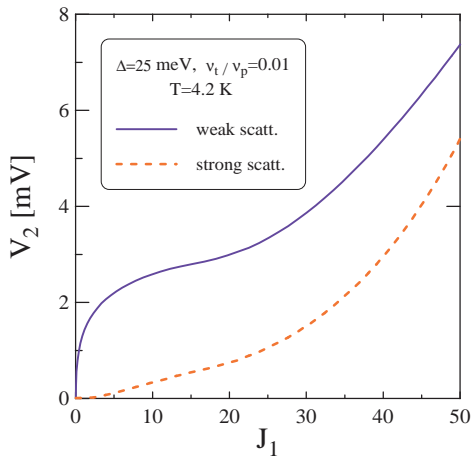,height=6cm}
\caption{Voltage measured at the second electrode as function of the
injection current
through the first electrode.}
\end{center}
\end{figure}

\subsection{Double-mesa experiment}

In this experiment which is described in detail in Ref.\onlinecite{Rother03} two
small mesas are structured close to each other on top of a common base mesa
and contacted with separate gold electrodes. Through the first mesa a
variable current $I_1$ is injected while at the second mesa the voltage $V_2$
is measured for fixed current $I_2$. Normally the voltage $V_2$ is
independent of the current $I_1$ as long as all junctions in the
base mesa are in the superconducting state.
In some cases, however, a small additional voltage $\Delta V_2(I_1)$ is
observed. This happens, if the lowest junctions in the two mesas are in the
resistive state (see Fig.\,7b) and generate a charge-imbalance on the first common
superconducting layer of the base mesa, which then depends on both currents.
\begin{figure}
\begin{minipage}{6.5cm}
\epsfig{figure=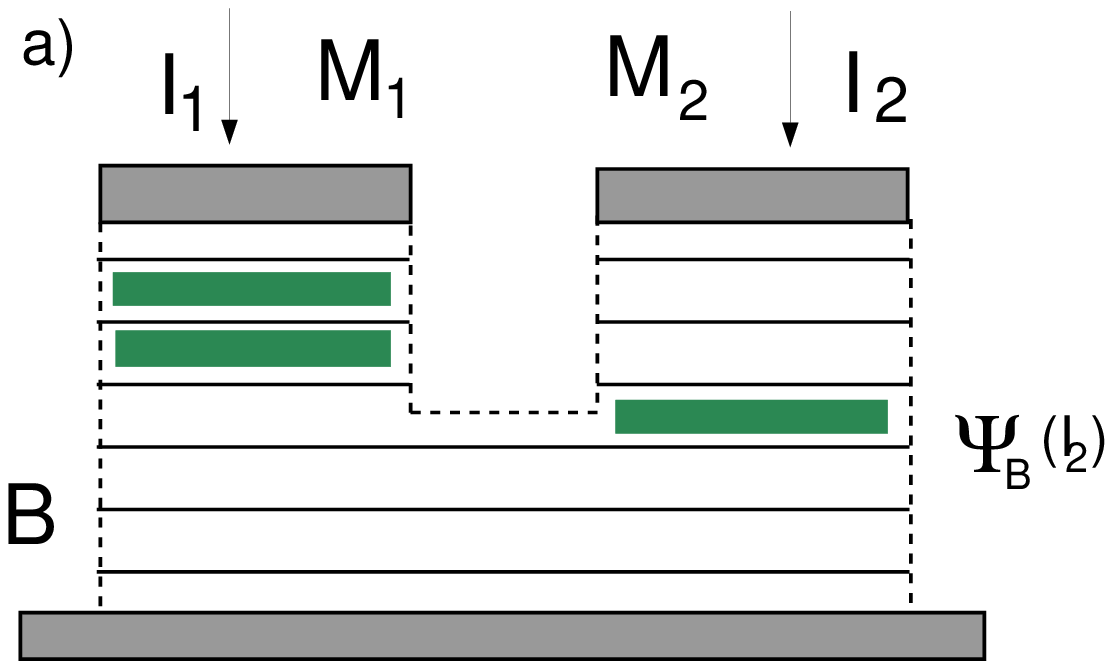,height=3.5cm}
\epsfig{figure=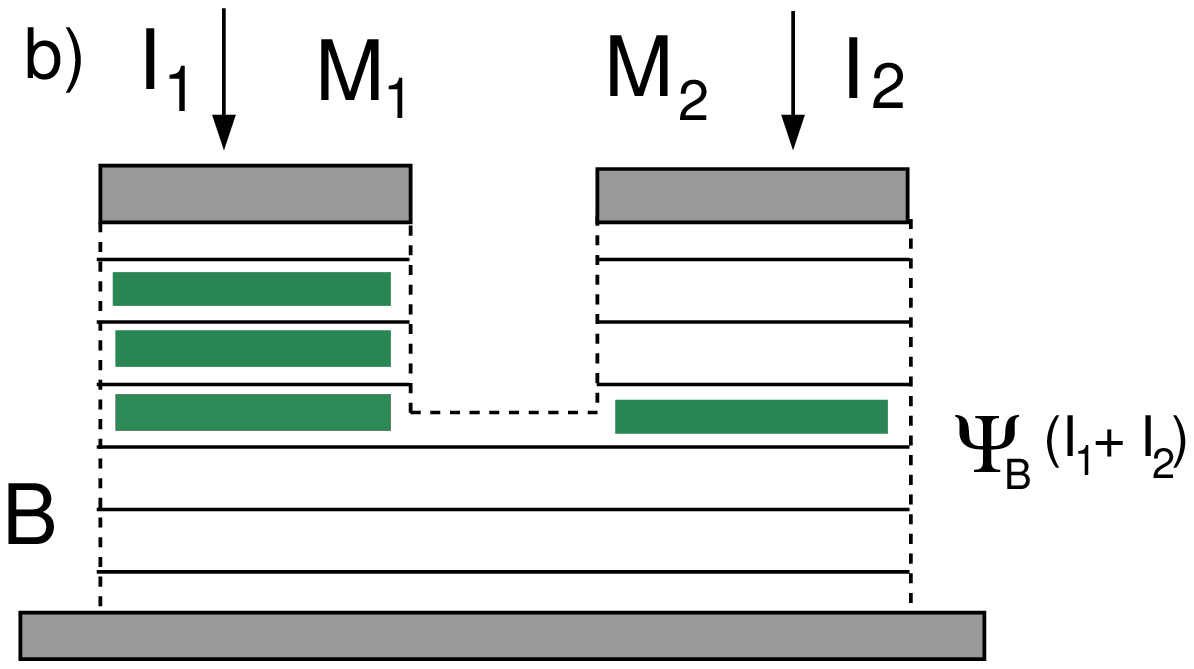,height=3.5cm}
\end{minipage}
\caption{Position of resistive barriers in the double-mesa experiment}
\end{figure}

If we want to apply the microscopic theory to this situation  we have to
calculate the charge-imbalance on a superconducting layer in contact with
another superconducting layer with the barrier being in the resistive state.
Furthermore we have to add to the  tunnelling current the average Josephson
current. In systems with a large McCumber parameter this contribution is
small. We will neglect it in the following.

In a first approximation we may
treat the  junctions between the two mesas and the base mesa as if
they were normal electrodes. In this case the only modification is the finite
current through the second mesa. The charge imbalance potential generated on
the first superconducting layer of the base mesa is
\begin{equation}
\Psi_B= \frac{F_1}{F} \psi(\frac{I_1}{F_1}) + \frac{F_2}{F}\psi(\frac{I_2}{F_2})
.
\end{equation}
The current density through the second mesa, which is kept constant, is
\begin{equation}
j_2=j_0(V_2) - \sigma_0 \Psi_B ,
\end{equation}
where $j_0(V)$ is the current-voltage function (\ref{current3}).
The voltage shift $\Delta V_2(I_1) =V_2(I_1)-\bar
V_2$, is the difference between the voltage measured at the second
mesa for fixed current $I_2$, when the last barriers in the first and second mesa
are in the resistive
state (Fig.\,7b), and the constant voltage $\bar V_2$, when only the last barrier of the
second mesa
is in the resistive state (Fig.\,7a).
Expanding $j_0(V)$ around the voltage $\bar V_2$ as
$j_0(\bar V_2 + \Delta V_2)=j_0(\bar V_2)+ \Delta V_2 \sigma(\bar V_2)$ we
obtain:
\begin{equation}
\Delta V_2(I_1) = \frac{\sigma_0}{\sigma(\bar V_2)}\frac{F_1}{F}
\psi(\frac{I_1}{F_1}) , \end{equation}
where the function $\psi(j)$ is shown in Fig.\,5b.
\begin{figure}
\begin{center}
\epsfxsize=0.7\hsize
\epsfig{figure=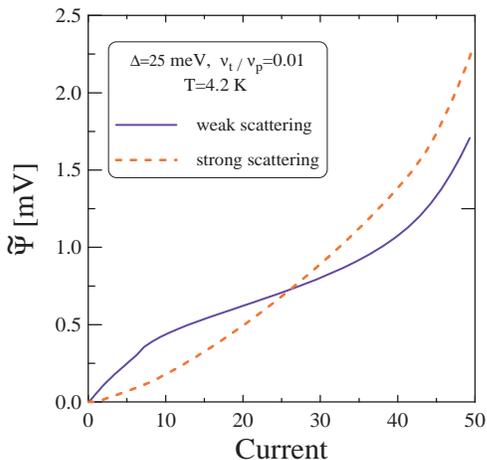,height=6cm}
\caption{Modified charge imbalance potential generated on a superconducting layer
by resistive junction as function of the tunnelling current.}
\end{center}
\end{figure}

This approximation can be improved if we take into account  that the last
layers in the small mesas are in the superconducting state. Then we have to
include the frequency dependent density of states $u'(\epsilon)$ of these
layers both in the calculation of the quasi-particle current densities
and in the charge imbalance function $\alpha(\epsilon)$.
Let us denote the modified quasi-particle current by $\tilde
j_0(V)$ and the modified charge-imbalance potential function by $\tilde
\psi(j)$ (explicite formulas are given in the Appendix),
 then the voltage shift $\Delta V_2(I_1)$ measured at the second mesa is
given by
\begin{equation}
\Delta V_2(I_1) = \frac{\sigma_0}{\tilde\sigma(V_2)}\frac{F_1}{F}
\tilde \psi(\frac{I_1}{F_1}) ,
\end{equation}
where $\tilde \sigma(V)=d\tilde j_0(V)/dV)$. Its current dependence is
proportional to the function $\tilde \psi(j)$, which is shown in Fig.\,8.
The shape of the functions $\psi(j)$, Fig.\,5b and $\tilde \psi(j)$, Fig.\,8
is similar. The curve obtained in the weak scattering limit (solid line)
also has great similarity
with the experimental results.\cite{Rother03} The parameters used in our
calculations are typical for these materials. We did not perform a
fit to the experimental data, since this would require an additional
parameter for the Josephson coupling. Of course all the curves end at the
critical current.

\subsection{Influence of charge-imbalance on the current-voltage curves}

The same formalism can be applied to calculate the influence of
charge-imbalance on the current-voltage curves in a stack of Josephson junctions. For
instance, for a stack of junctions  with one junction in the resistive state inside the
stack (not adjoining the normal electrodes) the current is given by
\begin{equation}
j(V)= \tilde j_0(V) - 2\sigma_0\tilde \Psi(V) ,
\end{equation}
where $V$ is the voltage drop across this barrier. The factor 2
comes from the charge-imbalance potential created on the two  superconducting
layers adjoining the barrier in the resistive state.

The corresponding
voltage shift for a given current $j$  is then
\begin{equation}
\Delta V= 2 \frac{\sigma_0}{\tilde \sigma(V)}\tilde \psi(j) ,
\end{equation}
where $V$ is the voltage for a
single junction. This shift can be measured
directly as difference between the voltage of two isolated junctions in the
resistive state and two neighboring resistive junctions in the stack. This generalises
our results   obtained with the phenomenological theory \cite{Ryndyk02} using an ohmic
quasi-particle IV-curve. Note that this (very small)  voltage
shift in the IV-curves does not depend on the parameter $a$ (called $\alpha$ in
Refs.\onlinecite{Koyama96, Ryndyk02} describing charge fluctuations of the superconducting condensate.
These will be of importance in dynamical effects like the Josephson
plasma resonance and in optical experiments. \cite{Helm02}

\section{Summary}

In this paper we have developed a comprehensive microscopic theory for stationary
nonequilibrium effects in  intrinsic Josephson systems starting
from a non-equilibrium Green function theory for layered d-wave
superconductors. We investigated the charge-imbalance generated on
a superconducting layer by current injection and  derived results for the
charge-imbalance distribution function and the nonequilibrium
quasi-particle current between superconducting layers. The theory  uses
basic nonequlibrium concepts developed earlier for bulk superconductors and
is applied here to  layered d-wave superconductors forming a stack of Josephson
junctions. Specific for
layered superconductors with small tunneling rate between the layers is the
confinement of charge-imbalance on single superconducting layers. Specific for
d-wave superconductors with vanishing gap is the relaxtion of charge-imbalance
due to elastic impurity scattering, which is the dominant relaxation  mechanism at
low energies. In  distinction to an earlier semi-phenomenological theory by the
authors we considered here the energy dependence of the charge-imbalance
distribution function and its  relaxation, leading to non-linear
current-voltage realtions.

We applied the theory to the calculation of nonequilibrium effects in
current injection experiments with 4 contacts. In particular, we
calculated the voltage  between a normal electrode and a
superconducting layer for zero current as function of the current through a
second electrode. This voltage measures directly the charge imbalance
potential generated on the superconducting layer. We then applied the theory
to recent double-mesa experiments. Thus we were able to
explain the non-linear dependence of the voltage measured at one mesa on the
current through the second mesa.

The same formalism can also be applied to calculate the influence of
charge imbalance on the  current-voltage curves in a stack of Josephson
junctions, which should be observable as difference in the voltage between different
configurations of a given number of resistive junctions in the stack. We note
that the  shift of the chemical potential of the condensate leads do a
redistribution of the voltage between different superconducting layers, but
this has no influence on the total current-voltage curves. Charge oscillations in the
condensate and the resulting coupling between the layers will however  be
important for dynamic effects like the dispersion of the   longitudinal
Josephson plasma resonance and in some optical experiments. These dynamic
effects will be investigated in a forthcoming publication.

\acknowledgments{
This work has been supported by a grant from the German Science Foundation.
We would like to thank P. M\"uller and R. Kleiner and their groups for a very fruitful
cooperation on the investigation of Josephson effects in layered
superconductors.}

\appendix
\section{Definition of non-equilibrium Green functions for layered superconductors}
\label{A1}

Following the standard definitions \cite{Larkin77,Larkin86} for non-equilibrium
Green functions for superconductors we define larger and
lesser Green functions for layers $n$ and $n'$ ($\bar k=-k$) in Nambu space
by
\begin{align}
&\hat G_{nn'}^>(\vec k, t_1;\vec k',t_2)= \\
& -i\begin{pmatrix} \langle
c_{nk\uparrow}(t_1) c^\dagger_{n'k'\uparrow}(t_2)\rangle &
\langle c_{nk\uparrow}(t_1) c_{n'\bar k'\downarrow}(t_2)\rangle \cr
-\langle
c^\dagger_{n\bar k\downarrow}(t_1)c^\dagger_{n'k'\uparrow}(t_2)\rangle &
-\langle c^\dagger_{n\bar k\downarrow}(t_1) c_{n'\bar k'\downarrow}(t_2)\rangle
\end{pmatrix}, \nonumber
\end{align}
\begin{align}
&\hat G_{nn'}^<(\vec k, t_1;\vec k',t_2)= \\
& +i\begin{pmatrix} \langle
c^\dagger_{n'k'\uparrow}(t_2) c_{nk\uparrow}(t_1)\rangle &
\langle  c_{n'\bar k'\downarrow}(t_2)c_{nk\uparrow}(t_1)\rangle \cr
-\langle
c^\dagger_{n'k'\uparrow}(t_2)c^\dagger_{n\bar k\downarrow}(t_1)\rangle &
-\langle  c_{n'\bar k'\downarrow}(t_2)c^\dagger_{n\bar k\downarrow}(t_1)\rangle
\end{pmatrix}, \nonumber
 \end{align}
from which we obtain the retarded advanced and Keldysh function by
\begin{align}
\hat G^R &=\Theta(t_1-t_2) (\hat G^> - \hat G^<), \nonumber \\
\hat G^A &=-\Theta(t_2-t_1) (\hat G^> - \hat G^<), \nonumber \\
\hat G^K &= \hat G^> + \hat G^<.
\end{align}
For the different components in Nambu space the
following notation is commonly used:
\begin{equation}
\hat G=\begin{pmatrix} G & F \cr - F^+ &\bar G\end{pmatrix}.
\end{equation}
For the average diagonal Green functions with $n'=n$ only one index and one
k-vector will be used. For these functions we introduce a Fourier transform
with respect to the time-difference $\tau:=t_1-t_2$ keeping the central
time $t:=(t_1+t_2)/2$ fixed:
\begin{equation}
\hat G(\vec k,t,\epsilon)= \int d\tau \hat G(\vec k,t+\frac{\tau}{2}, t-\frac{\tau}{2})
e^{i\epsilon \tau}.
\end{equation}
Finally we introduce the quasi-classical approximation by integrating over
the energy $\xi_k=\epsilon_k-\mu$ keeping the direction $\hat k$ of the
momentum fixed:
\begin{equation}
\hat g(\hat k,t,\epsilon)=\begin{pmatrix}g & f \cr -f^+ & \bar
g\end{pmatrix}=\frac{i}{\pi} \int d\xi_k \hat G(\vec
k,t,\epsilon).
\end{equation}

\section{Derivation of  kinetic equations}
\label{A2}

In our model the equations of motion for
the Keldysh Green function on layer $n$  in Nambu space are given by:
\begin{align}
i\frac{\partial }{\partial t_1}\tau_3 \hat G^K_n(\vec k,t_1,t_2) &- \hat H(\vec
k,t_1) \hat G^K_n(\vec k,t_1,t_2) \nonumber \\
&- \bigl\{\hat \Sigma^R \hat G^K
+\Sigma^K G^A\bigr\} = 0, \nonumber \\
-i\frac{\partial }{\partial t_2} \hat G^K_n(\vec k,t_1,t_2)\tau_3
&- \hat G^K_n(\vec k,t_1,t_2)\hat H(\vec
k,t_2) \nonumber \\& - \bigl\{\hat G^R \hat \Sigma^K +  \hat G^K
\Sigma^A\bigr\} = 0,
\end{align}
where
\begin{equation}
\hat H(\vec k,t)= -e \phi_n(t) + \xi_k - \hat \Delta_k(t),
\end{equation}
with the electical scalar potential $\phi_n(t)$, the charge of the electron
$(-e)$, and the order parameter matrix
\begin{equation}
\hat \Delta_k(t)=\begin{pmatrix} 0 & \Delta_k(t) \cr -\Delta_k^*(t) & 0
\end{pmatrix}.
\end{equation}
Here $\Delta_k(t)= \Delta_k e^{i\chi_n(t)}$ has a time-dependent phase. The
constant amplitude $\Delta_k$ of the order parameter is equal on each layer
and has d-wave symmetry. In order to eliminate the time-dependent phase in
the order parameter we make a gauge transformation
$c_{nk}(t)=\tilde c_{n,k}(t)\exp(i\chi_n(t))$. After this gauge transformation
the new Green function $\tilde G$ fulfills an equation of motion with
$\hat H(\vec k,t)= -e \Phi_n(t) + \xi_k - \hat \Delta_k$, the gauge invariant
scalar potential $\Phi_n(t)= \phi_n(t)- \hbar\dot
\chi_n(t)/2e$, and a real time-independent order parameter.

The symbol $\{\hat
\Sigma \hat G\}$ denotes a convolution in time space of the self-energy and the
Green function:
\begin{equation}
\{AB\}(t_1,t_2)=\int dt_3A(t_1,t_3)
B(t_3,t_2).
\end{equation}
The self-energy, which will be discussed in detail later,
contains random impurity scattering within the layer and tunnelling to the
neighboring layers. The latter will treated in second order in the tunnelling
matrix element.

A kinetic equation is obtained subtracting the two equations, introducing
a central time $t:=(t_1+t_2)/2$ and taking the Fourier transform with respect
to the time-difference $\tau:=t_1-t_2$:
\begin{eqnarray}
&&(\frac{i}{2}\frac{\partial}{\partial t}+\epsilon) \tau_3 \hat G^K +
(\frac{i}{2}\frac{\partial}{\partial t}-\epsilon)  \hat G^K \tau_3
 - \bigl\{\hat H \hat G^K - \hat G^K \hat H\bigr\} \nonumber \\
&&= \bigl\{\hat \Sigma^R \hat G^K + \hat \Sigma^K \hat G^A - \hat G^R \hat
\Sigma^K - \hat G^K \hat \Sigma^A\bigr\}.
\end{eqnarray}
In this equation the kinetic energy $\xi_k$
drops out and we can perform the integration over $\xi_k$ keeping the
direction $\hat k$ of the momentum fixed. We then obtain the kinetic
equation for the quasi-classical Green functions
\begin{align}
(\frac{i}{2}\frac{\partial}{ \partial t} + \epsilon)  \tau_3 \hat
g^K &+(\frac{i}{ 2}\frac{\partial}{ \partial t} - \epsilon) \hat
g^K \tau_3 \nonumber \\
& -\bigl\{\hat h(\hat k,t)  \hat g^K-  \hat g^K\hat
h(\hat k,t)\bigr\} \nonumber \\
 =\bigl\{\hat \sigma^R \hat g^K+& \hat
\sigma^K\hat g^A -\hat g^R\hat \sigma^K - \hat g^K \hat\sigma^A\bigr\}
\label{eqq}
\end{align}
with
\begin{align}
\hat h(\hat k,t) &= -e\Phi_n(t) - \hat \Delta(\hat k), \nonumber \\
\hat \Delta(\hat
k) &= \begin{pmatrix} 0 & \Delta(\hat k) \\
-\Delta(\hat k) & 0 \end{pmatrix}.
\end{align}
The curly brackets denote a
convolution in time and frequency space:
\begin{equation}
\{AB\}(t,\epsilon)=e^{i(\partial_t^A\partial_\epsilon^B-\partial_\epsilon^A\partial_t^B)/2}
A(t,\epsilon)B(t,\epsilon).
\end{equation}
The equation (\ref{eqq}) is the starting point
of our calculations.

\section{Coupling between two superconducting layers}

\noindent
a) {\textit Tunnelling between two superconducting layers with  the barrier in the
superconducting state}

Here we have $\Omega_{n,n'}=0$ but a finite  constant phase difference
$\varphi:= \gamma^0_{n,n'}$, and we obtain for the tunnelling self-energy
\begin{align}
\hat \sigma_t^R(\hat k,\epsilon)_{n,n'}= &- i
\langle \nu_t{g}^R(\epsilon)\rangle' \tau_3
+  \sin\varphi\langle \nu_t
{f}^R(\epsilon)\rangle'  \tau_1 \nonumber \\
&+ \cos \varphi \langle \nu_t
{f}^R(\epsilon)\rangle'  \tau_2,
\end{align}
\begin{align}
\hat \sigma_t^K(\hat k,\epsilon)= &- 2i
\langle \nu_t u'(\epsilon)\beta'(\epsilon)\rangle' \tau_3
- 2i \langle \nu_t u'(\epsilon)\alpha'(\epsilon)\rangle' \tau_0 \nonumber\\
&+ 2 \sin\varphi \langle
\nu_t v'(\epsilon)\beta'(\epsilon)\rangle'  \tau_1 \nonumber\\
&+2 \cos \varphi\langle
\nu_t v'(\epsilon)\beta'(\epsilon)\rangle'  \tau_2 \nonumber\\
&+ 2 \cos\varphi \langle
\nu_t w'(\epsilon)\alpha'(\epsilon)\rangle'  \tau_1 \nonumber\\
&-2 \sin\varphi\langle
\nu_t w'(\epsilon)\alpha'(\epsilon)\rangle'  \tau_2.
\end{align}
For the contribution from tunnelling to the scattering term in the kinetic
equation we then find:
\begin{align}
I^{ts}_{11}-
I^{ts}_{22} &= -8i \langle \nu_t u'(\epsilon)
u(\epsilon)\alpha(\epsilon)\rangle' +4i \langle \nu_t
u'(\epsilon)\alpha'(\epsilon)  u(\epsilon)\rangle', \nonumber \\
J^{ts}_{12}-
J^{ts}_{21}&= 8i\langle \nu_t
u'(\epsilon)\alpha'(\epsilon) v(\epsilon)\rangle'.
\end{align}
Note that  all the contributions depending on the phase difference $\varphi$ drop out. The
remaining terms vanish if we neglect terms of order
$\nu_t\alpha$ which are of higher order in the tunnelling probability.

\medskip
\noindent
b) {\textit Tunnelling between two superconducting layers with the barrier in the
resistive state.}

In this case also the anomalous Green functions contribute to the dc current.
As this contribution involves the coherent part of the tunnelling matrix
element and will be small in the limit of large McCumber parameters, we
neglect it in the following. However, we have to take into account
the frequency dependent density of states
$u(\epsilon), u'(\epsilon)$ on both superconducting layers. We want to apply
the theory to calculate the charge-imbalance $\alpha$ on a superconducting
layer $n$ which is coupled on one side to a superconducting layer $n'$ in a
stack  with barriers in the resistive state and on the other side to a
superconducting layer with the barrier in the superconducting state.
As the layer $n'$ is between two barriers in the resistive state with equal
quasi-particle current we can neglect the charge-imbalance $\alpha'$ on this
layer. Then we obtain the contribution to the scatterng  term from
tunnelling between $n$ and $n'$:
\begin{align}
I^{tn}_{11}-
I^{tn}_{22}
=-4i\langle
\nu_t&(u'(\epsilon-v)-u'(\epsilon+v))
u(\epsilon)\beta(\epsilon)\rangle'\nonumber
\\ +4i\langle \nu_t(u'(\epsilon-v)&\beta'(\epsilon-v ) -
u'(\epsilon +v) \beta'( \epsilon+v))u(\epsilon)\rangle' \nonumber \\
-4i\langle \nu_t (u'(\epsilon-v)
&+u'(\epsilon+v))u(\epsilon)\alpha(\epsilon)\rangle', \nonumber \\
J^{tn}_{12}-J^{tn}_{21} = 4i \bigl\langle \nu_t& v(\epsilon)) \bigl(
u'(\epsilon-v) \beta'(\epsilon-v) \nonumber \\
 - &u'(\epsilon+v) \beta'(\epsilon+v)\bigr)
\bigr\rangle'.
\end{align}
For the charge-imbalance function $\alpha(\epsilon)$ on the superconducting layer
$n$ we then find:
\begin{align}
2\nu_p\bigl[u(\epsilon)\langle u(\epsilon)\alpha(\epsilon)\rangle
&-\langle u(\epsilon)\rangle u(\epsilon) \alpha(\epsilon)\bigr] \nonumber \\
&-2\nu_p\frac{\Delta
v(\epsilon)}
{\tilde \epsilon} \langle u(\epsilon)\alpha(\epsilon)\rangle \nonumber \\
= - \bigl(u(\epsilon)- \frac{\Delta}{ \tilde
\epsilon}v(\epsilon)\bigr)
&\bigl\langle \nu_t(\hat k,\hat k') \bigl[
u'(\epsilon-v)(\beta'(\epsilon-v) - \beta(\epsilon)) \nonumber \\
+ u'(\epsilon+v)(\beta(\epsilon)& - \beta'(\epsilon+v))\bigr] \bigr\rangle'.
\end{align}

The formulas for the current densities obtained from Eq.(\ref{current2})
now read in (symmetrised form)
\begin{align}
 j^\beta(V)  &= \frac{e}{2\hbar}  N(0) \nu_t
\int_{-\infty}^{+\infty} d\epsilon \langle u(\epsilon)\rangle
\langle
u'(\epsilon+eV)  \\
&+u'(\epsilon-eV)\rangle(\beta_0(\epsilon+eV)-\beta_0(\epsilon-eV)) =:\tilde
j_0(V), \nonumber \\
j^\alpha(V) &= -\frac{e}{\hbar}  N(0) \nu_t \int_{-\infty}^{+\infty} d\epsilon \tilde
\alpha(\epsilon) \langle u'(\epsilon+eV) \nonumber \\
&+u'(\epsilon-eV)\rangle =: -\sigma_0
\tilde \Psi(V), \end{align}
with the charge imbalance function $\tilde \alpha(\epsilon)=\langle
u(\epsilon)\alpha(\epsilon)\rangle$ given by
\begin{align}
\tilde \alpha(\epsilon)= \frac{\nu_t \langle
R(\epsilon)\rangle}{2\nu(\epsilon)}&
\langle u'(\epsilon+eV)+u'(\epsilon-eV) \nonumber \\
&\rangle\bigl(\beta_0(\epsilon+eV)-\beta_0(\epsilon-eV)\bigr) .
\end{align}
The current equation for $j^\alpha(V)$ defines a modified charge-imbalance
potential $\tilde \Psi(V)$, from which we obtain the function $\tilde
\psi(j)=\tilde \Psi(V)$ using the current-voltage relation $j=\tilde j_0(V)$.
This function is shown in Fig.\,8.

\section{Charge imbalance of quasi-particles}

In order to make contact with the traditional theory of charge imbalance let
us summarise the basic definitions and concepts of charge-imbalance as
introduced by Tinkham and Clarke
\cite{Tinkham72,Tinkham,Clarke72,Clarke86}.

In the BCS theory the total charge (in units of the electron
charge, factor 2 from spin)  is
\begin{equation}
Q=2\sum_k v^2_k +
(u^2_k-v^2_k)f_k =:Q^c+Q^* ,
\end{equation}
where $u^2_k=\frac{1}{2}(1+\xi_k/E_k)$,
$v^2_k=(1+\xi_k/E_k)/2$ are the usual coherence factors,
$\xi_k=\epsilon_k-\mu$, $E_k=\sqrt{\xi_k^2+\Delta_k^2}$ is the quasi-particle
excitation energy. The first term is the condensate charge $Q^c$, the second
term is the quasi-particle charge $Q^*$, $f_k$ is the quasi-particle
distribution function. In  equilibrium $f_k=1/(\exp(E_k/T)+1)$ and hence $Q^*$
vanishes (for particle-hole symmetry).

A nonequilibrium
state can be described by a shift $\delta \mu$ of the chemical potential
and a change of the distribution function $f_k$. A shift of the chemical
potential  leads to a shift of the excitation
energy in k-space, $\xi_k \to \epsilon_k - \mu + \delta \mu$ and hence to a
change of the condensate charge by:
\begin{equation}
\delta Q^c= 2N(0) \delta \mu ,
\end{equation}
where $N(0)$ is the  density of states for one spin
direction at the Fermi energy.

The quasi-particle charge can also be written as
\begin{equation}
Q^*= 2 \sum_k q_k  f_k =\sum_k q_k(f_{k_>}-f_{k_<}),
\end{equation}
where $q_k=\xi_k/E_k$ can be considered as charge of a quasi-particle
with momentum $\vec k$. For each
quasi-particle state with  momentum $\vec k=\vec k_>$ and $\xi_{k_>}>0$
(electron-like quasi-particle) exists a quasi-particle state with momentum
$\vec k=\vec k_<$ and $\xi_{k_<}=-\xi_{k_>}<0$ (hole-like quasi-particle) with
the same excitation energy $E_k>0$ and direction $\hat k$. Thus $Q^*$
depends on the difference in the number of electron- and hole-like
quasi-particles. While $ f_{k_>}-f_{k_<}$ describes charge-imbalance, the
combination $f_{k_>}+f_{k_<}$, for instance, enters the self-consistency
equation for the gap $\Delta_k$.

These distribution functions are introduced here for
well-defined quasi-particles with infinite life-time. In a microscopic
theory based on nonequilibrium Green functions these are replaced here by the
frequency-dependent distribution functions $\alpha(\epsilon)$ and
$\beta(\epsilon)$. In our case we find the following correspondence for
$\epsilon>0$:
\begin{equation}
\beta(\epsilon=E_k)= 1-f_{k_>} - f_{k_<},
\end{equation}
\begin{equation}
\alpha(\epsilon=E_k)= - q_k(f_{k_>}- f_{k_<}).
\end{equation}
Extended to the whole frequency range $\beta(\epsilon)$ becomes an odd
function, $\alpha(\epsilon)$ is an even function.


\begin{thebibliography}{10}

\bibitem{Kleiner92} R. Kleiner, F. Steinmeyer, G. Kunkel, and P. M\"uller,
Phys. Rev. Lett. {\bf 68}, 2394 (1992)

\bibitem{Kleiner94}
R. Kleiner and P. M\"{u}ller, Phys. Rev. B {\bf 49},  1327  (1994);
M\"uller P in: {\it Festk\"orperprobleme/Advances in Solid State Physics}, Vol 34,
ed. by Helbig R (Vieweg, Braunschweig), 1 (1994)

\bibitem{Yurgens96}
A. Yurgens, D. Winkler, N. V. Zavaritsky, T. Claeson, Phys. Rev. {\bf 53}, R8887 (1996)

\bibitem{Yurgens00}
A. A. Yurgens, Superconductor Science and Technology {\bf 13},  R85  (2000)

\bibitem{Doh00}
Y. J. Doh, J. Kim, K. T. Kim, and H. J. Lee, Phys. Rev. B {\bf 61},  R3834
  (2000)

\bibitem{Rother03}
S. Rother, Y. Koval, P. M\"uller, R. Kleiner, D. A. Ryndyk, J. Keller, Phys. Rev. B{\bf 67},
024510 (2003)

\bibitem{Ryndyk02} D. A. Ryndyk, J. Keller, C. Helm, J. Phys.: Condens Matter {\bf
14}, 815 (2002)

\bibitem{Ryndyk98}
D. A. Ryndyk, Phys. Rev. Lett. {\bf 80},  3376  (1998)

\bibitem{Ryndyk99}
D. A. Ryndyk, JETP {\bf 89},  975  (1999) [Zh. Eksp. Teor. Fiz. {\bf 116}, 1798
  (1999)]

\bibitem{Clarke72}
J. Clarke, Phys. Rev. Letters {\bf 28}, 1363 (1972); M. Tinkham, J. Clarke, Phys. Rev.
Letters {\bf 28}, 1366 (1972)

\bibitem{Tinkham}
M. Tinkham, {\it Introduction to Superconductivity}, Second Edition, Chapter 11,
McGraw-Hill, 1996

\bibitem{Langenberg}
D. N. Langenberg, A. I. Larkin, editors {\it Nonequilibrium Superconductivity},
North-Holland, 1986

\bibitem{Clarke79}
J. Clarke, U. Eckern, A. Schmid, G. Sch\"on, and M. Tinkham, Phys. Rev. B {\bf 20}, 3933
(1979)

\bibitem{Pethick79}
C. J. Pethick, H. Smith, Ann.Phys. (N.Y.) {\bf 119}, 133 (1979)

\bibitem{Schmid75}
A. Schmid, G. Sch\"on, J. Low Temp. Phys. {\bf 20}, 207 (1975)

\bibitem{Schoen86}
G. Sch\"on, in  {\it Nonequilibrium Superconductivity}, D. N. Langenberg, A. I. Larkin,
editors,  North-Holland, 1986

\bibitem{Larkin77}
A.I. Larkin, Yu. N. Ovchinnikov, Soviet.Phys. JETP {\bf 46}, 155 (1977)

\bibitem{Larkin86}
A. I. Larkin, Yu. N. Ovchinnikov, in {\it Nonequilibrium Superconductivity}, D. N. Langenberg,
A. I. Larkin, editors,  North-Holland, 1986

\bibitem{Artemenko80}
S. N. Artemenko, Zh. Eksp. Teor. Fiz. {\bf 79}, 162 (1980) [Sov. Phys. JETP {\bf 520},
81 (1980)]

\bibitem{Graf95} M. J. Graf, M. Palumbo, D. Rainer, and J. A. Sauls, Phys. Rev. B {\bf 52}, 10588
(1995)

\bibitem{Koyama96}
T. Koyama and M. Tachiki, Phys. Rev. B {\bf 54},  16183  (1996)

\bibitem{B96}
L. N. Bulaevskii, D. Dominguez, M. Maley, A. Bishop, and B. Ivlev, Phys. Rev. B
  {\bf 53},  14601  (1996)

\bibitem{Artemenko97}
S. Artemenko and A. Kobelkov, Phys. Rev. Lett. {\bf 78},  3551  (1997)

\bibitem{Preis98}
C. Preis, C. Helm, J. Keller, A. Sergeev, and R. Kleiner, in {\it Superconducting
Superlattices II: Native and Artificial}. I. Bozovic and D. Pavona, editors, Proceedings
of SPIE Volume {\bf 3480},  236  (1998)

\bibitem{Shafranjuk99}
S. E. Shafranjuk, M. Tachiki, Phys. Rev. B {\bf 59}, 14087 (1999)

\bibitem{Helm01}
C. Helm, J. Keller, C. Preis, and A. Sergeev, Physica C {\bf 362}, 43 (2001)

\bibitem{Helm00} C. Helm, C. Preis, C. Walter,
J. Keller, Phys. Rev. B {\bf 62}, 6002 (2000)

\bibitem{Helm02}
C. Helm, L. N. Bulaevskii, E. M. Chudnovsky, M. P. Maley, Phys. Rev. Lett. {\bf
89}, 057003 (2002)

\bibitem{Bulaevskii02}
L. N. Bulaevskii, C. Helm, A. R. Bishop, M. P. Maley, Europhys Lett. {\bf
58}, 057003  (2002)

\bibitem{Matsumoto99} H. Matsumoto, S. Sakamoto, F. Wajima, T. Koyama, M. Machida,
Phys. Rev. B {\bf 60}, 3666 (1999)

\bibitem{Tinkham72}
M. Tinkham, Phys. Rev. B{\bf 6}, 1747 (1972)

\bibitem{Clarke86}
J. Clarke in {\it Nonequilibrium
Superconductivity}, D. N. Langenberg, A. I. Larkin, editors,  North-Holland,
1986

\bibitem{Schmid81}
A. Schmid, in {\it Nonequilibrium Superconductivity, Phonons, and Kapitza
Boundaries} (Proc. NATO ASI), K. A. Gray (ed.) Plenum Press, New York (1981), p.
423.

\bibitem{Wang01} H. B. Wang, P. H. Wu, T. Yamashita, Phys. Rev. Lett. {\bf 87},
107002 (2001)

\end{thebibliography}
\end{document}